\documentclass[review]{elsarticle}

\usepackage{lineno,hyperref}
\usepackage[T1]{fontenc} 
\usepackage{amsmath,amsfonts,amsbsy,amssymb}
\usepackage{mathabx} 
\usepackage{amssymb} 
\usepackage{amsmath}
\usepackage{amsthm}	
\usepackage{mathrsfs}
\usepackage[nolist]{acronym} 
\usepackage{tabularx} 
\usepackage{multirow}
\usepackage{wasysym}
\usepackage{float}
\usepackage{color}

\begin{document}

\begin{frontmatter}

\title{Performance of Non-orthogonal Multiple Access under Finite Blocklength}

\author[mymainaddress]{Endrit Dosit}

\author[mysecondaryaddress]{Mohammad Shehab\corref{mycorrespondingauthor}}
\cortext[mycorrespondingauthor]{Corresponding author}
\ead{mohammad.shehab@oulu.fi}
\author[mysecondaryaddress]{Hirley Alves}
\author[mysecondaryaddress]{Matti Latva-aho}

\address[mymainaddress]{Department of Electrical and Computer Engineering, University of Toronto, Canada.}
\address[mysecondaryaddress]{Centre for Wireless Communications (CWC), University of Oulu, Finland}

\begin{abstract}
In this paper, we present a finite-block-length comparison between the orthogonal multiple access (OMA) scheme and the non-orthogonal multiple access (NOMA) for the uplink channel. First, we consider the Gaussian channel, and derive the closed form expressions for the rate and outage probability. Then, we extend our results to the quasi-static Rayleigh fading channel. Our analysis is based on the recent results on the characterization of the maximum coding rate at finite block-length and finite block-error probability. The overall system throughput is evaluated as a function of the number of information bits, channel uses and power. We find what would be the respective values of these different parameters that would enable throughput maximization. Furthermore, we analyze the system performance in terms of reliability and throughput when applying the type-I ARQ protocol with limited number of retransmissions. The throughput and outage probability are evaluated for different blocklengths and number of information bits. Our analysis reveals that there is a trade-off between reliability and throughput in the ARQ. While increasing the number of retransmissions boosts reliability by minimizing the probability of reception error, it results in more delay which decreases the throughput. Nevertheless, the results show that NOMA always outperforms OMA in terms of throughput, reliability and latency regardless of the users priority or the number of retransmissions in both Gaussian and fading channels. 
\end{abstract}

\begin{keyword}
non-orthogonal multiple access (NOMA), finite blocklength, ultra reliable communication (URC).
\end{keyword}

\end{frontmatter}

\nolinenumbers

\section{Introduction} \label{Introduction}
Driven by the market demands for extra services, the fifth generation of mobile communication is expected to provide seamless connectivity enabling Machine Type Communication (MTC) \cite{MTC} and the Internet of Things (IoT) \cite{IoT}. The IoT is considered as the next revolution in mobile cellular systems. It interconnects "things" (such as machines, sensors, smart meters) and facilitates autonomous data exchange between them bringing connectivity to anything that can benefit from internet connection \cite{ericsson,Networks}. Depending on their functionality, these devices will have different requirements and constraints. They might have limitations on hardware, energy efficiency, reliability, latency or even a combination of those. Most of the results that are present so far in the wireless communication theory are based on the Shannon capacity \cite{Shannon_paper}. This is an asymptotic metric, which corresponds to the highest data rate that we can communicate, given that we want to maintain a certain level of reliability. It is asymptotic in the sense that if the transmission is done at a certain rate, which is lower than the Shannon capacity, then it would be possible to achieve arbitrarily low error probability only by using sufficiently long packets. This metric has been very successful so far because of large delay constrains ( e.g. 10 ms or more), and very large packet lengths (e.g. 10s of thousands channel uses) \cite{Devassy2014}. Another classical metric which is used is the outage capacity, which is the extension of the Shannon capacity to non-ergodic channels \cite{Biglieri1998}. 

Meanwhile, short packet communication has become a mandatory solution in order to satisfy extremely low latency as envisioned for real time applications and emerging technologies such as e-health and road safety. Despite their small size, these packets need to be decoded at the intended receiver with very high reliability and very low latency, which of course depends on the system specifications. In the finite block-length regime, where the packets are short, the situation changes drastically. Since one of their fundamental assumptions (very large packet length) does not hold anymore, these metrics become a poor benchmark \cite{Durisi_1,iswcs18}. From an information theoretic point of view, communication on short packets does not comply to traditional metrics such as Shannon capacity \cite{paper11}. Alternatively, low latency communication is subject to the finite blocklength capacity model, where the length of metadata is of comparable size to the length of actual data. \cite {paper2,paper5,paper14}. In this context, the evaluation of the maximum achievable rate $R(n,\epsilon)$ as a function of block-length $n$ and error probability $\epsilon$ has gathered much attention. It is defined as the largest coding rate for which there exists an encoder/decoder pair $(f_n,g_n)$ of packet length $n$, whose packet error probability does not exceed $\epsilon$. During these last few years, a lot of progress has been done in developing very tight non-asymptotic upper and lower bounds of $R(n,\epsilon)$ for both Additive White Gaussian Noise (AWGN) and fading channels in \cite{Polyanskiy2010,Yang}.

Recently, the study of finite block-length communication, has brought significant progress in the field of wireless communications with application to Ultra-Reliable Low-Latency Communications (URLLC) \cite{Popovski2014}. URLLC has emerged to provide solutions for reliable and low latency transmissions in wireless systems. The design of URLLC systems imposes strict quality of service (QoS) constraints to fulfill very low latency in the order of milliseconds with expected reliability of higher than 99.9$\%$ \cite{NokiacMTC2016,eucnc}. In \cite{latency}, Schulz et al. discussed the reliability requirements for different IoT applications. According to their study, latency bounds range from 1 ms in factory automation to 100 ms in road safety, while the packet loss rate constraints range from $10^{-9}$ in printing machines to $10^{-3}$ for traffic efficiency. Such requirements are far more stringent than the ones in the current long term evolution (LTE) standards \cite{Johan}. In this context, in \cite{me} the authors develop optimal power allocation strategy for type-I Automatic Repeat Request (ARQ) protocol, which enable communications at any target outage probability in finite blocklength. Furthermore, they show that the proposed strategy maximizes the overall system throughput in the ultra reliable region. Similarily, in \cite{Makki2014} the authors focus on minimizing the outage probability given a certain average power constraint for type-I ARQ protocol. The scheme proposed, is valid only when the maximum number of transmissions is set to two. In addition, the obtained results are valid only for a single user scenario. 

Nowadays Orthogonal Multiple Access (OMA) schemes is the multiple access scheme mostly used in communication systems. Therein, the time and frequency resources are shared between users. Another approach is the Non-Orthogonal Multiple Access (NOMA) presented in \cite{Poor, Higuchi2015}. The key feature of NOMA is to serve multiple users at the same time/frequency/ code, but with different power levels, which yields a significant spectral efficiency gain over conventional orthogonal MA \cite{Poor}. In \cite{Higuchi2015}, the authors introduce another multiple access scheme, in which they exploit the power domain. In this scenario, all the resources are shared between users during the transmission. Then, their packets are demultiplexed in the base station (BS) using Successive Interference Cancellation (SIC) and following a certain pre-determined order.  In their paper, it is shown that the rates of each of the users are increased if NOMA is utilized. The authors of \cite{SIC} discussed the NOMA strategy for massive IoT where they derived the system stability requirements. They found out that the optimal strategy is to divide the whole bandwidth into a few sub-bands when there is no delay constraint and to utilize it as one sub-band when a delay constraint is imposed.

Similar to \cite{Higuchi2015}, in our system the highest priority is given to a certain user, and thus its transmitted packet is decoded last. In our analysis, we assume to have multiuser uplink setup, where all of the users will communicate to the same access point. Depending on their system requirements, these users will have different priorities. Therefore, various questions arise: ``How will all these users transmit their information in a channel whose resources they share?'', ``How will the BS process the information, given that the reliability requirements are met?'', ``How would it be possible to meet these requirements,while spending minimal amount of resources?'', and finally "How to achieve all this in a delay limited system?".

In this paper, we analyze two different multiple access approaches, the orthogonal and non-orthogonal scheme to determine which one performs better in finite block-length. Not many works have been presented in the field of designing and analyzing multiple access schemes in this regime. In \cite{Devassy2014}, the authors analyze the scenario of having several uncoordinated users who transmit short coded packets, using frequency hopping and an automatic repeat request (ARQ) protocol. However, the interference is simply treated as noise. While in \cite{Tandon2013} the analysis is done for specific combinations of modulation and coding schemes. Part of the work in this journal was presented in \cite{balkan} where we characterized the throughput of orthogonal and non-orthogonal scheme in both AWGN and fading channels. 

Herein, we extend the previous work by analyzing the network throughput for each user when applying the type-I ARQ protocol. According to \cite{5GNR}, HARQ is considered to be a part of the 5G New Radio standard. When compared to other reliability enhancement techniques such as Chase Combining Hybrid Automatic Repeat Request (CCHARQ) \cite{EUCNC_endri} and relaying, type-I ARQ has lower complexity and latency. Therefore, we choose type-I ARQ as a simple method to improve reliability in ULLRC networks at minimum complexity and latency cost. In type-I ARQ protocol, each user is allowed to retransmit its packet if it receives a NACK feedback from the BS, which means that the packet was not successfully decoded. Each user possesses $M$ trials to transmit a single packet, where $M$ is limited in order to suppress extra latency. The results will show that although applying ARQ reduces the throughput, it offers higher reliability in terms of lower outage probabilities specially for the NOMA scheme. Moreover, the latency penalty is less punishing for NOMA when compared to OMA.

The rest of the paper is organized as follows: in  Section \ref {sc:system model}, we introduce the system model for NOMA and OMA in AWGN and quasi-static Rayleigh fading channels. Communication in finite blocklength is characterized in Section \ref{sc:FB}, where we define the achievable rate and then derive outage expressions. Next, Section \ref{sc:ARQ} analyzes the throughput of type-I ARQ protocol with limited number of retransmissions. We present the numerical results and comparisons between OMA and NOMA schemes in terms of reliability, throughput, and latency in Section \ref{Numerical results}. Finally, Section \ref{con} concludes the paper. On the last page, Table 1 includes the important abbreviations and symbols that will appear throughout the paper.

\begin{table}[t]
	\caption{List of abbreviations and symbols.}
	\begin{tabular}{cccc}
		\hline
		ARQ & Automatic Repeat reQeust \\		
		NOMA & Non-Orthogonal Multiple Access\\
		QoS & Quality Of Service \\
		SINR & Signal-to-Interference-plus-Noise Ratio \\		
		URLLC & Ultra Reliable Low Latency Communication		
   \vspace{5mm} \\

		$C(\cdot)$ & Shannon capacity \\
		$\mathbb{E}[ \ ]$ & expectation of \\
		$\mathcal {K}$ & successfully transmitted information bits at the $m^{th}$ round \\ 		
		$N$ & number of nodes \\
		$\Pr[\cdot]$ & probability of \\
		$Q(x)$ & Gaussian Q-function \\
		$Q^{-1}(x)$ & inverse Gaussian Q-function \\
		$R(n,\epsilon)$ & achievable rate \\
		$\mathcal{T}$ & total channel uses at the $m^{th}$ round \\ 		
		$V(\cdot)$ & channel dispersion \vspace{5mm} \\

		$e$ & exponential Euler's number \\
		$h_i$ & fading coefficient of user $i$ \\
		$k$ & number of transmitted information bits \\
		$n$ & blocklength \\
		$w$ & additive while Gaussian noise vector \\
		$x_i$ & transmitted signal vector of user $i$ \\
		$y$ & received signal vector \\
		$z$ & fading random variable \vspace{6mm} \\
		
        $\beta$ & user 1 quota of channel uses in OMA \\
		$\epsilon$ & error probability \\
		$\eta$ & throughput \\
		$\rho$ & signal to noise ratio \\				
\hline
	\end{tabular}
\end{table}

\section{System model} 
\label{sc:system model}

We start by investigating the OMA and NOMA schemes for both AWGN and fading channels. In the fading channel model, we assume that users are communicating in a quasi-static Rayleigh fading channel with coherence bandwidth $B_c$ and coherence time $T_c$. For the OMA scenario, these coherence intervals are assigned to users depending on their priority. This implies that, $\beta$ portion of the available channel uses will be assigned to one user, and the remaining $1-\beta$ channel uses will be assigned to the other user. On the other hand, for the NOMA case the users can utilize all the resources whenever needed. Furthermore, we assume that the user transmissions will be uncoordinated. This implies that there will be interference between them. Each user in the system will transmit his packet, which will consist of $n$ symbols with duration $n/B_c < T_c$. This guarantees that the channel will remain constant for the entire duration of the packet, while our analysis accounts for the short blocklength effect. 

The received siganl vector $y \in C^n$ consists of the coded packets $x_i \in C^n$, which were transmitted during the coherence interval. Thus, we have:
\begin{equation}
\label{eq:a}
y = \sum_{i=1}^{s} h_ix_i + w,
\end{equation}
where $h_i$ is the fading coefficient for user $i$ which is assumed to be quasi-static with Rayleigh distribution and independent identically distributed from one packet duration to the other. $x_i$ is the transmitted packet of user $i$, and $ w_i$ corresponds to the additive noise vector, whose entries are modeled as zero-mean circularly symmetric Gaussian random variables with unit variance. Further in \eqref{eq:a}, the index $s$ spans over the entire set of interfering users.

In this model, we assume that the receiver is aware of the priority of the users. For the OMA scenario, this implies that the receiver knows which part of the available resources are allocated to each user. While, in the NOMA scenario, it only knows the decoding order to perform SIC \cite{Ekram}. For the fading channel, we assume that the receiver has channel state information (CSI). 

\begin{figure}[!htb] 
	\centering
	\includegraphics[clip, trim= 8cm 7cm 10cm 4cm, width=1\textwidth]{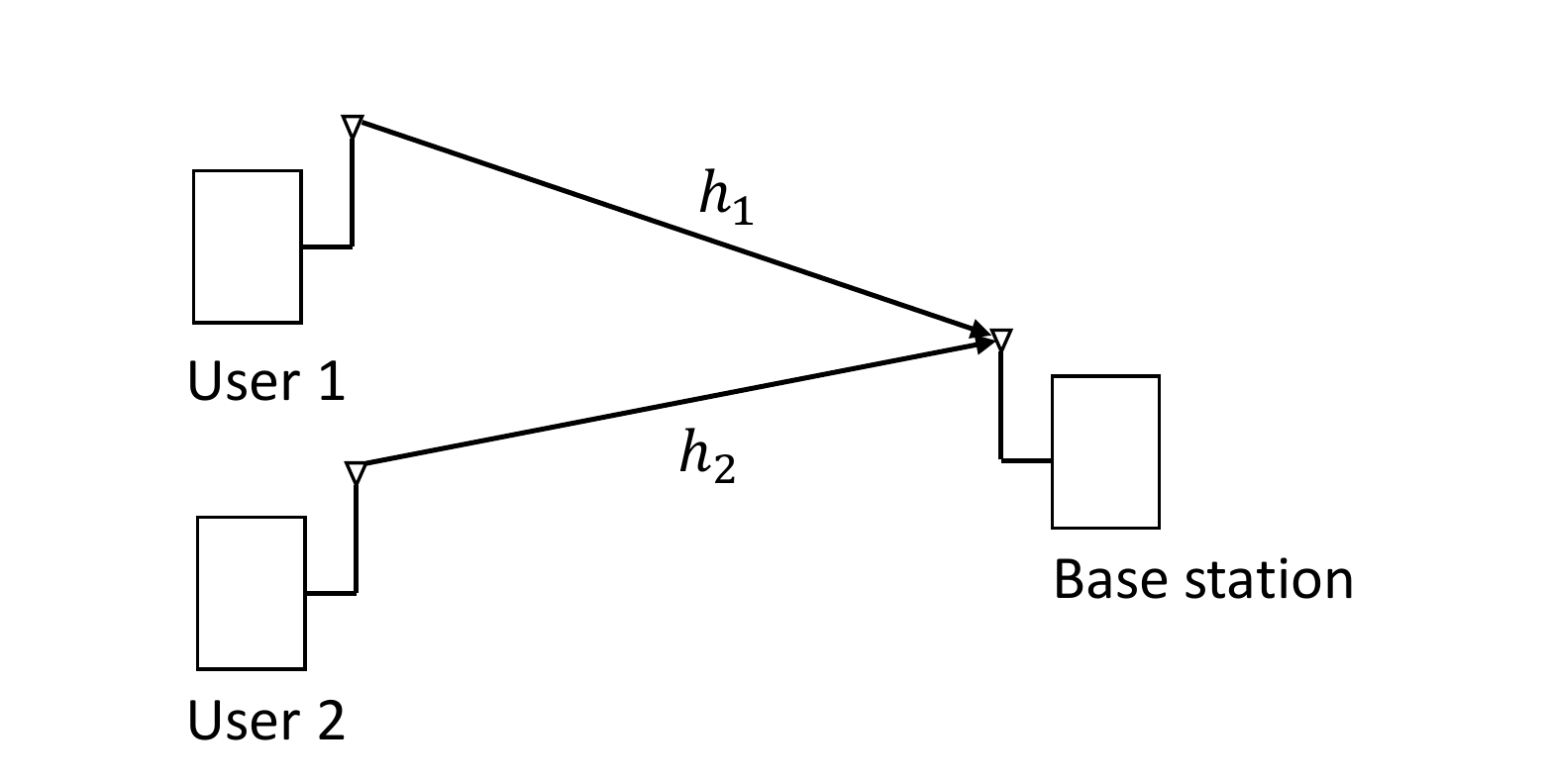}
	\caption{Uplink channel model with two users.}
	\label{fig:2_us}
	\vspace{-0mm}
\end{figure}

The number of information bits in each transmitted packet is $k$; thus all users will have the same payload. Furthermore, we suppose that these information bits are mapped to the same number of channel uses $n$ in order to minimize the transmission delay. For simplicity, we assume that there are only two users in our system as in \cite{2users}. The uplink model is depicted in Fig. \ref{fig:2_us} where 2 users transmit short data packets to the BS through interference channel. Each user transmits its data across different channel conditions where the CSI information is assumed to known at the BS. Moreover, We assume that User 1 has higher priority so the BS performs SIC to eliminate the interference resulting from User 2 on User 1 and thus, User 1 suffers no interference.

\section{Maximum coding rate and outage at finite block-length}
\label{sc:FB}

Herein, we characterize the individual rates and outage probabilities of the users in our system for both NOMA and OMA scenarios. First, we analyze the AWGN channel and then, we proceed with the Rayleigh fading channel as in \cite{Turkish}.

\subsection{AWGN channel} 
\label {a}
To model the AWGN channel, we assume that the channel gains in \eqref{eq:a} have unity gain. Thus, we would obtain the received signal as
\begin{align}
y = x_1 + x_2 + w.
\end{align}
From \cite{Polyanskiy2010}, we compute the maximum achievable rate for a single user as:
\begin{align}
\label{R}
R(n,\epsilon)= C(\rho) - \sqrt {\frac{V(\rho)}{n}}  Q^{-1} (\epsilon)+ \operatorname{O}\left( \frac{\log_2n}{2n}\right),
\end{align}
where $\rho$ represents the signal-to-noise (SNR) ratio, $Q(t)=\int_{t}^{\infty}\frac{1}{\sqrt{2 \pi}}e^{\frac{-s^2}{2}} ds$ is the Gaussian Q-function, and $Q^{-1} (t)$ represents its inverse. C($\rho$) is the channel capacity calculated as
\begin{align}
\label{eq:3}
C(\rho) = \log_2 (1+\rho).
\end{align}
Further in \eqref{R}, V($\rho$) denotes the channel dispersion, which can be found from \eqref{eq:4}
\begin{align}
\label{eq:4}
V(\rho) = \left(1-\frac{1}{(1+\rho)^2}\right)  \log_2^2 (e),
\end{align}
and $\epsilon$ is the packet error probability given in \eqref{eq:5}:
\begin{align}
\label{eq:5}
\epsilon = Q \left( \frac{C(\rho)+0.5\log_2 (n) -k}{\sqrt{n  V (\rho)}}\right),
\end{align}
where $k$ is the number of transmitted information bits. The throughput of the system for a certain user can be found as
\begin{align}
\label{throughput}
\eta=\frac{k}{n} \left(1-\epsilon\right)
\end{align}

Notice that in \eqref{R} the remainder terms $\frac{\log_2 n}{2n}$ and $\operatorname{O}(1)$ vanish as $n$ increases. Thus, we neglect these terms from \eqref{R} and compute the individual user rates for OMA scenario as follows:
\begin{equation}
\label{eq:6}
\begin{split}
R_1&=\left(  C(\rho_1)-\sqrt{\frac{  V(\rho_1)}{ \beta n}}Q^{-1}(\epsilon)\right), \\
R_2\!&=\!\left( C(\rho_2)- \sqrt{\frac{V(\rho_2)}{(1- \beta)n}}Q^{-1}(\epsilon)\right),
\end{split}
\end{equation}
where $\beta$ refers to the amount of resources that we allocate to the first user. 

A very important metric used to evaluate the performance of a system is the outage probability of the user transmissions. For our system, it is derived from \eqref{eq:5} as 
\begin{align}
\label{eq:7}
\begin{split}
\epsilon_1 &= Q \left(\frac{ C(\rho_1) - k}{ \sqrt{n \beta V(\rho_1)}}\right), \\
\epsilon_2 &= Q \left(\frac{ C(\rho_2) - k}{ \sqrt{n ( 1- \beta )  V(\rho_2)}}\right).
\end{split}
\end{align}

In the NOMA scenario and after applying SIC at the BS , the SINR of the users changes as
\begin{align}
\label{eq:8}
\rho_1 &= P_1 \\
\rho_2 &= \frac{P_2}{1 + P_1}
\end{align}
The rates and outage probability can be computed from \eqref{eq:6} and \eqref{eq:7} by setting $\beta=1$ for the first user and $\beta=0$ for the second user. 

\subsection{Quasi-static fading channel}
\label{b}
Now we turn to the quasi-static fading channel case where the channel coefficient remains constant during one transmission. The channel gains $h$, change independently between transmissions with exponentially distributed probability density function (pdf). Thus, the received signal would be \cite{NOMA_arxiv}:
\begin{align}
\label{eq:9}
y=h_1x_1+h_2x_2+w.
\end{align}

From \cite{Yang4}, the maximum achievable rate for fading channels is approximated as:
\begin{align}
\label{eq:10}
R(n,\epsilon)=C_\epsilon + \operatorname{O}\left(\frac{\log{n}}{n}\right),
\end{align}
where $\operatorname{O}(\frac{\log n}{n})$ is remainder term and $n$ is the number of channel uses. Notice that as $n$ increases, $\operatorname{O}(\frac{\log n}{n}) \approx 0$. Finally, in \eqref{eq:10} $C_\epsilon$ denotes the outage capacity, which is computed from
\begin{align}
\label{eq:11}
C_\epsilon= \sup  \left\{R: \Pr[\log_2 (1+\rho Z) < R] < \epsilon\right\},
\end{align}
where $Z \thicksim \mathrm{Exp}(-z)$ and represents the squared envelope of the channel coefficients according to \cite{alouini}. Let $\rho$ represent the S(I)NR of the received signal, and let $N_0=1$.  Also, $R$ refers to the ratio between the number of information bits that we have to transmit $k$, and the number of channel uses $n$, thus $R=\frac{k}{n}$. The outage probability $\epsilon$, for the quasi static fading channel can be found as \cite{Polyanskiy2010}
\begin{align}
\label{eq:12}
\epsilon= \mathbb{E} \left\{ Q \left( \frac{\left(C(\rho z) - k\right)}{\sqrt{n V(\rho z )}]}\right) \right\}.
\end{align}

In the OMA scenario, the outage probability can be found from \eqref{eq:12}. We recall that user $1$ utilizes $\beta n$ channel uses, while user $2$ utilizes the remaining $(1-\beta) n$ channel uses. Notice that \eqref{eq:12} is not a closed form solution. However, it can be well-approximated by linearizing the $Q$ function as in \cite{Makki2014,Q_linear}. For this purpose we write $Q (f(z))$ as
\begin{align}
\label{eq:19}
Q (f(z)) = W(z) = \left\{
\begin{array}{ll}
1 & if \quad z \leq \sigma_i,  \\
0.5 \!- \frac{b_i}{\sqrt{2\pi}} (z - \theta_i)  & if \quad \sigma_i <\! z <\! \delta_i,  \\
1 & if \quad z \geq \delta_i,
\end{array}
\right.
\end{align}
where
\begin{align}
\label{eq:15}
\theta_1&= \frac{e^\frac{k}{\beta n}-1}{\rho_1},\\
\label{ecr}
\theta_2&= \frac{e^\frac{k}{(1-\beta) n}-1}{\rho_2},\\
\label{eq:16}
b_1&=\sqrt{\frac{\beta n \rho_1^2}{e^\frac{2k}{\beta n}-1}},\\
\label{ecp}
b_2&=\sqrt{\frac{(1-\beta )n \rho_1^2}{e^\frac{2k}{(1-\beta)n}-1}},\\
\label{eq:17}
\sigma_i&=\theta_i- \sqrt{\frac{\pi}{2b_i^2}},\\
\label{eq:18}
\delta_i&=\theta_i+ \sqrt{\frac{\pi}{2b_i^2}}.
\end{align}

Notice that for the NOMA case, we set $\beta=1$ in \eqref{eq:15} and \eqref{eq:16}. While, in \eqref{ecr} and \eqref{ecp} we set $\beta=0$ to guarantee full sharing of resources. Further, in \eqref{eq:17} and \eqref{eq:18}, $i \in \{1,2\}$ for user 1 and user 2, respectively.

Then, we write the integral form of $\epsilon$ as
\begin{align}
\label{eq:20}
\epsilon_i=\int_{0}^{\infty} W(z)  f_Z(z) \mathtt{d}z,
\end{align}
where $f_Z(z)$ corresponds to the probability density function of $Z$. After solving the integral for the OMA scenario, we obtain the outage expression in \eqref{eq:21}, which has the same form for both users
\begin{align}
\label{eq:21}
\epsilon_i=1-\frac{b_i}{\sqrt{2\pi}}e^{-\theta_i}\left(e^{\sqrt{\frac{\pi}{2b_i^2}}}-e^{-\sqrt{\frac{\pi}{2b_i^2}}}\right).
\end{align}

Next we characterize the case of NOMA with SIC. As mentioned in Section \ref{sc:system model}, we assume that user $1$ has the highest priority. This implies that its packets are decoded last, which results in an interference free decoding. Therefore, its outage probability can be computed from  \eqref{eq:21}. However, the situation changes in the case of the user who is decoded first. To facilitate calculations, we consider that the interference is much larger than noise, and thus the following approximation holds
\begin{align}
\label{eq:k}
\rho_2 (z)=\frac{P_2|h_2|^2}{P_1|h_1|^2}.
\end{align}
Therefore, the distribution of S(I)NR changes as a result of the presence of Rayleigh channel coefficients in both the numerator and denominator. To solve the integral presented in \eqref{eq:20}, we need to compute the pdf of $\rho_2 (z)$ which is given as
\begin{align}
\label{eq:j}
f_{\rho_2( z )} =\frac{P_1P_2}{(z P_1 + P_2)^2}.
\end{align}
Then, by substituting \eqref{eq:k} and \eqref{eq:j} in \eqref{eq:20}, we can compute the integral. Finally, the outage probability of this user is given by \eqref{y} on the top.
\begin{figure*}[!t]
	\begin{align} 
	\label{y}
	\epsilon_2  =& \frac{2\frac{b_2}{\sqrt{2\pi}}P_2^2(\delta_2-\sigma_2 + 2P_1^2\delta_2\sigma_2}{2(P_2+P_1\delta_2)(P_2+P_1\sigma_2)}  + \frac{\frac{b_2}{\sqrt{2\pi}}P_2\log(\frac{P_2+P_1\sigma_2}{P_2 + P_1\delta_2})}{P_1} \notag   \\ &+\frac{P_1P_2(\delta_2 + 2\frac{b_2}{\sqrt{2\pi}}\theta_2\delta_2 + \sigma_2 - 2\frac{b_2}{\sqrt{2\pi}}\theta_2\sigma_2}{2(P_2+P_1\delta_2)(P_2+P_1\sigma_2)}.
	\end{align}
	\hrule 
\end{figure*}

\section{ARQ and Throughput}\label{sc:ARQ}
As discussed in Section \ref{Introduction}, finite blocklength communication plays a significant role in URLLC. However, achieving the ultra low outage probabilities needed for URLLC with a single transmission is a cumbersome task. As will be shown is Section \ref{Numerical results}, when an open loop setup is utilized (one shot transmission), the outage probability for both the users is relatively high. To cope with this issue, we rely on schemes that exploit diversity.

In this context, one well established approach is the deployment of retransmission schemes. Specifically, in this section we focus on Automatic Repeat Request (ARQ) scheme \cite{ARQ}. Therein, each user retransmits the same packet in different transmission slots. The receiver fetches each of these packets, and checks if it can decode them. Once successful decoding of the packets is performed, a positive acknowledgment (ACK) packet is sent to the transmitter, who then moves on to sending new information. Assuming that each transmission faces different fading realizations, the outage probability after $M$ transmission rounds will be:
\begin{align}
\label{eq:outage}
\epsilon_M=\prod_{m=1}^M \epsilon_m \mathrm{,}
\end{align}
where $\epsilon_m$ is the outage probability of the $m^{th}$ transmission. The outage probability before the first transmission, $\epsilon_0=1$. 

We know that the utilization of schemes which are based on retransmissions causes the latency of a system to increase due to the presence of feedback, which is then reflected in the overall system throughput. In this section, we shall analyze the impact of type I ARQ scheme on the throughput. 

In the open loop setup, the throughput of each user can be found as:
\begin{align}
\label{eq:26}
\eta_i=\frac{k}{n}(1-\epsilon_i ),
\end{align}
where the outage probability $\epsilon_i$ of each user is given from (\ref{eq:21}). When the throughput is analyzed, it is important to take into account that each transmission is associated with a degradation of spectral efficiency. As a result of this degradation, we can say that the codeword rate in the $m^{th}$ ARQ round would be $R=\frac{k}{mn}$. The total number of channel uses in the $m^{th}$ round as the sum of transmissions time till reaching the $m^{th}$ round plus the feedback overhead can be computed from
\begin{align}
\label{eq:27}
\mathcal {T}=mn\sum_{i=1}^{m} \epsilon_{i-1}+D(m-1)\sum_{i=1}^{m-1} \epsilon_{i-1},
\end{align}
where $D$ is the feedback delay expressed in channel uses and $\epsilon_m$ is the packet drop probability Since transmissions are independent and outage probabilities are equal in all transmissions, the expression in \eqref{eq:outage} reduces to the product of the outage probabilities of each round given by $\epsilon_M=\epsilon_i^M$. It is clear that utilizing ARQ boosts reliability to the power $M$ order. Next we find the expected number of information bits that will be transmitted successfully without error after $M$ trails as:
\begin{align}
\label{eq:30}
\mathcal {K}=k(1-\epsilon_M).
\end{align}

Finally, assuming that a maximum number of $M$ retransmissions occurs, the throughput would be the ratio of successfully transmitted bits to the total number of channel uses given by
\begin{align}
\label{eq:31}
\eta_i=\frac{\mathcal {K}}{\mathcal {T}}&=\frac{k(1-\prod_{m=1}^M \epsilon_m)}{Mn\sum_{m=1}^{M} \epsilon_{m-1}+D(M-1)\sum_{m=1}^{M-1} \epsilon_{m-1}} \notag \\
&=\!\frac{k(1-\epsilon_i^M)}{Mn(1+(M-1)\epsilon_i)\!+\!D(M-1)(1+(M-1)\epsilon_i)} \notag \\
&=\frac{k(1-\epsilon_i^M)}{(1+(M-1)\epsilon_i)\left( Mn+D(M-1)\right)}.
\end{align}
Here $\epsilon_{m-1}$ refers to the outage probability up to the $m^{th}$ round. The main cost of this approach can be clearly seen from the equation above and is the throughput degradation due to the presence of feedback and transmission of multiple packets. The situation is a trade-off between reliability and delay. Since this throughput degradation becomes worse as we increase the number of retransmissions and we are also delay-limited in our system, we shall limit our analysis for the case when we have a maximum of 3 transmissions in our system.

\section{Numerical results}
\label{Numerical results}

In this section, we compare the performance of OMA and NOMA for $\beta=0.8$. Thus, 80$\%$ of the resources are allocated to the User 1 (the user with high priority), and 20$\%$ to the User 2. First, we start by the AWGN channel, then we proceed with fading case and ARQ protocol.

\subsection{AWGN channel}
\label{crap}
\begin{figure}[!htb] 
	\centering
	\includegraphics[clip, trim=1cm 0cm 0cm 0cm, width=1.08\columnwidth]{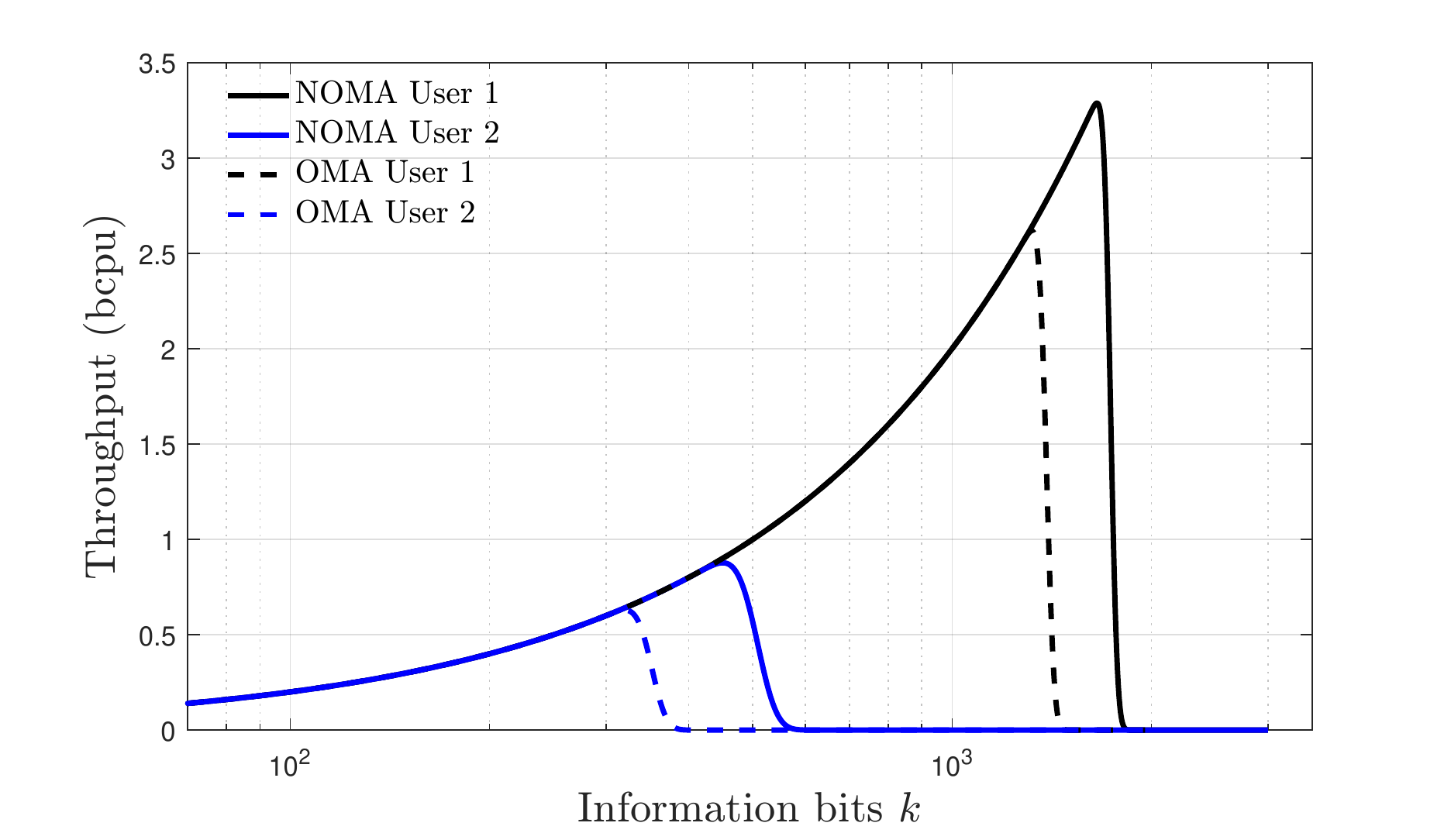}
	\caption{Throughput as a function of the number of information bits, considering $n= 500$ channel uses and $P_1 = P_2=10$ dB.}
	\label{fig:k_Th}
\end{figure}
For AWGN channel, we examine the throughput of each multiple access scheme. First, the throughput is analyzed as a function of the number of information bits $k$. For this purpose, we fix the number of channel uses to $n=500$. Furthermore, the transmitted power of users is set to $10$ dB. Since the noise has unit variance, this power value will directly correspond to the S(I)NR. In Fig.\ref{fig:k_Th}, we plot the throghput of NOMA and OMA as a function of $k$ for both users. We observe that NOMA scheme achieves higher throughput with respect to OMA. Apparantely, user 1 has greater throughput than user 2 because user 1 was assumed to have higher priority when applying the SIC. It is also noticed that different users have different values of $k$ for which this throughput is maximized. This result fully matches the system requirements. Since the users are communicating independently, their messages may have different sizes.

\begin{figure}[!htb] 
	\centering
	\includegraphics[clip, trim=1cm 0cm 0cm 0cm,width=1.08\columnwidth]{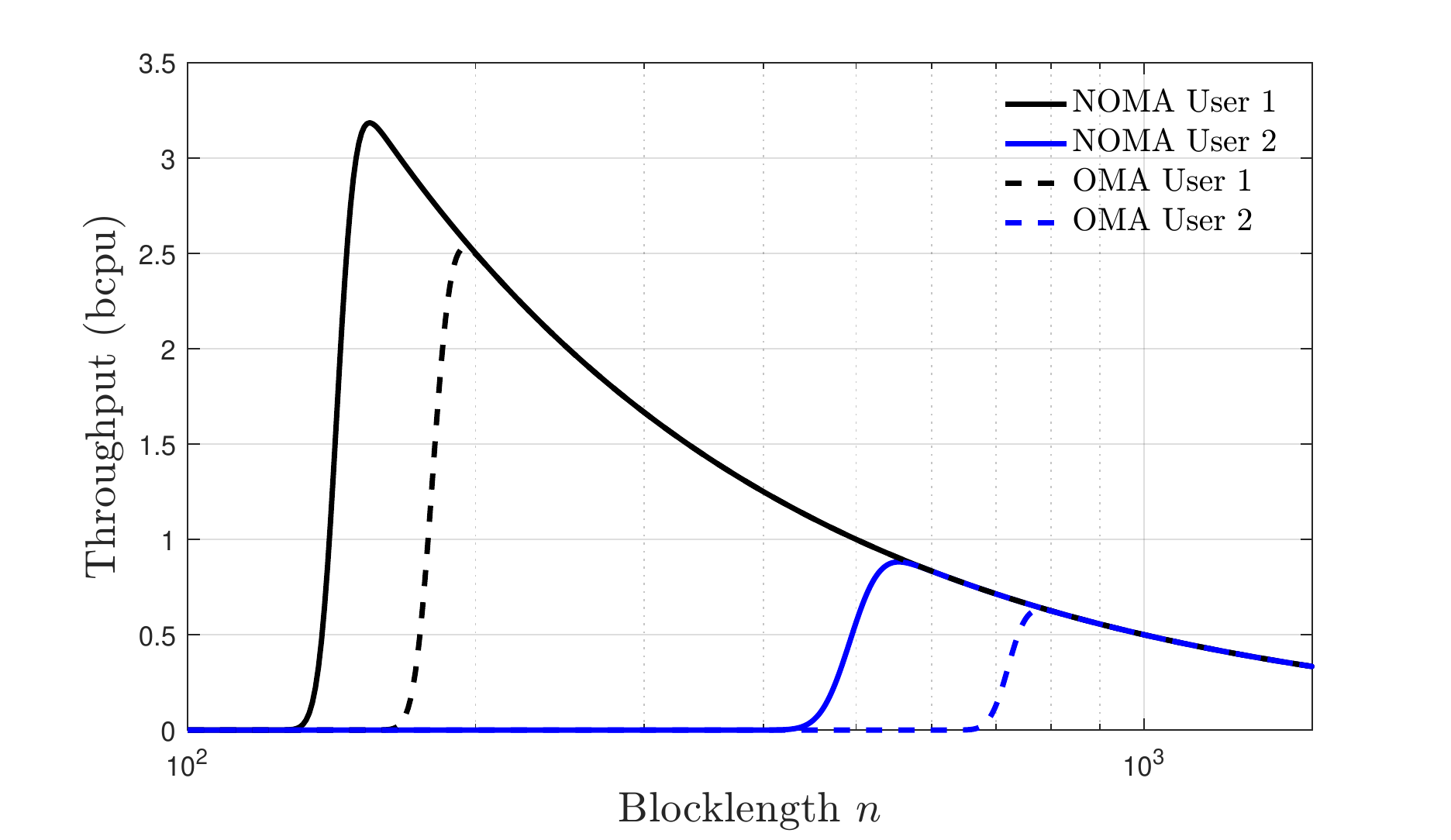}
	\caption{Throughput as a function of the number of channel uses, considering $k= 500$ bits, $P_1 = P_2=10$ dB.}
	\label{fig:n_Th}
\end{figure}

Next, we analyze the behavior of the throughput as a function of the channel uses. For this purpose, we fix $k=500$ bits and the transmitted power of both users in the system to $10$ dB. The results can be observed from Fig. \ref{fig:n_Th}. Notice that the approximations derived in Section. \ref{a} hold only for the block-lengths of size $n \geq 100$ channel uses \cite{Polyanskiy2010}. Again, NOMA scheme performs better than OMA. Furthermore, we notice that for different users, we have higher throughput for user 1 and different values of $n$ for which the throughput is maximized. Similarly as above, this result matches the intuition. 

\begin{figure}[!htb] 
	\centering
	\includegraphics[clip, trim=1cm 0cm 0cm 0cm,width=1.08\columnwidth]{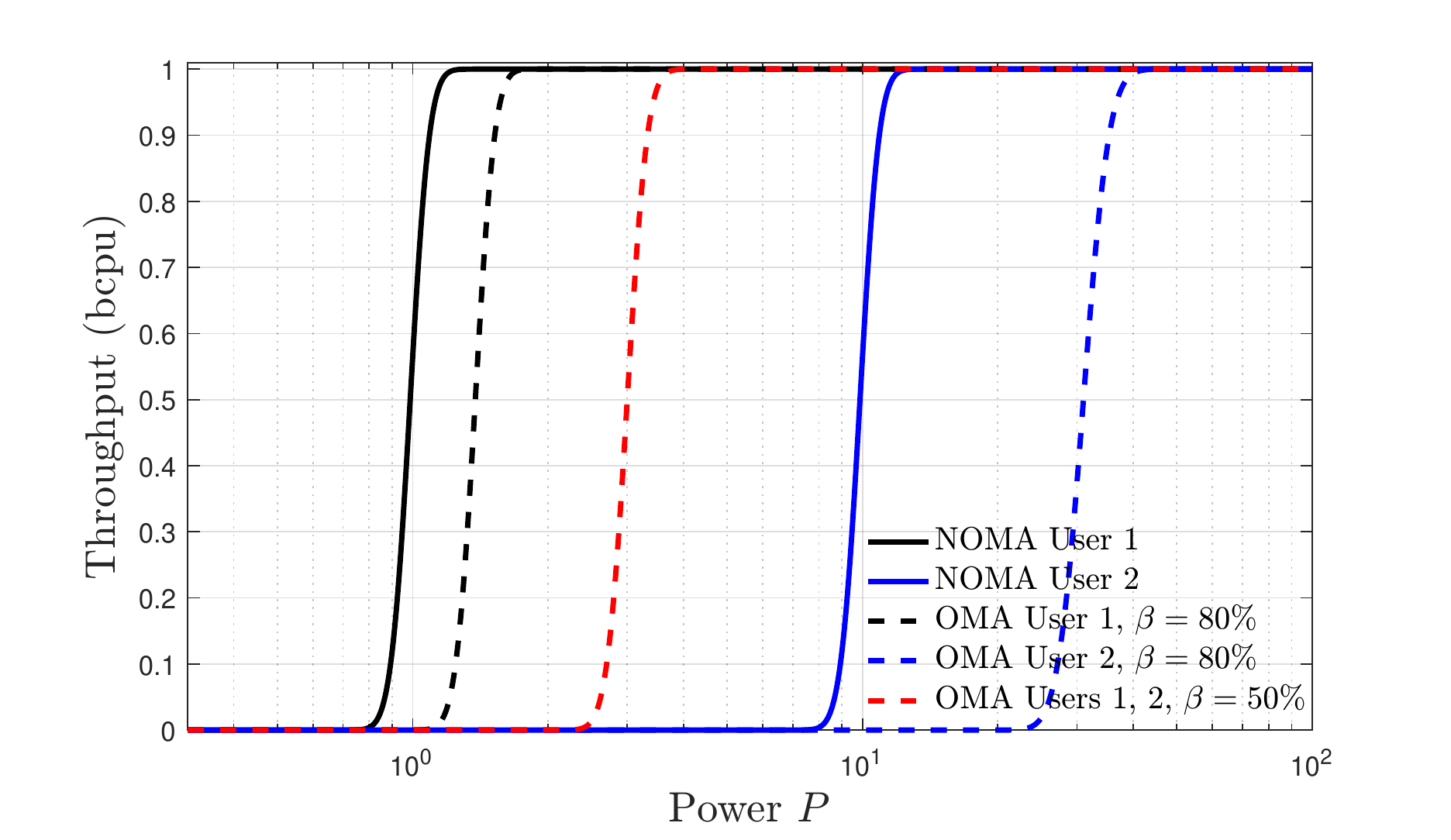}
	\caption{Throughput  as a function of transmitted power for $k=n=500$ for NOMA and OMA with different values of $\beta$}
	\label{fig:P_Th}
\end{figure}

Finally, for the AWGN channel we analyze the throughput as a function of transmission SNR for $k=n=500$. The result is shown in Fig. \ref{fig:P_Th}. To attain this plot, we fixed the channel coding rate at $R=1$ bpcu. Here, for the OMA scheme, both scenarios when highest priority is given to one user (i.e. $\beta=80\%$), and when both users are treated equally (i.e. $\beta=50\%$) are evaluated. We observe that the throughput is maximized at low SNR levels and then it saturates. This implies that, increasing the transmission power above a certain level will not result in an increase of throughput. Furthermore, we notice that if the utilized multiple access scheme is NOMA, the throughput is maximized at lower SNR levels than OMA for each of the users. However, when OMA scheme is utilized and equal resources to both users are assigned, we observe that for the second user we can achieve higher throughput with lower power expenditure. Nevertheless, for the first user the power savings that come from the implementation of NOMA are significantly higher. 

\subsection{Fading channel}
\begin{figure}[!htb] 
	\centering
	\includegraphics[clip, trim=1cm 0cm 0cm 0cm,width=1.08\columnwidth]{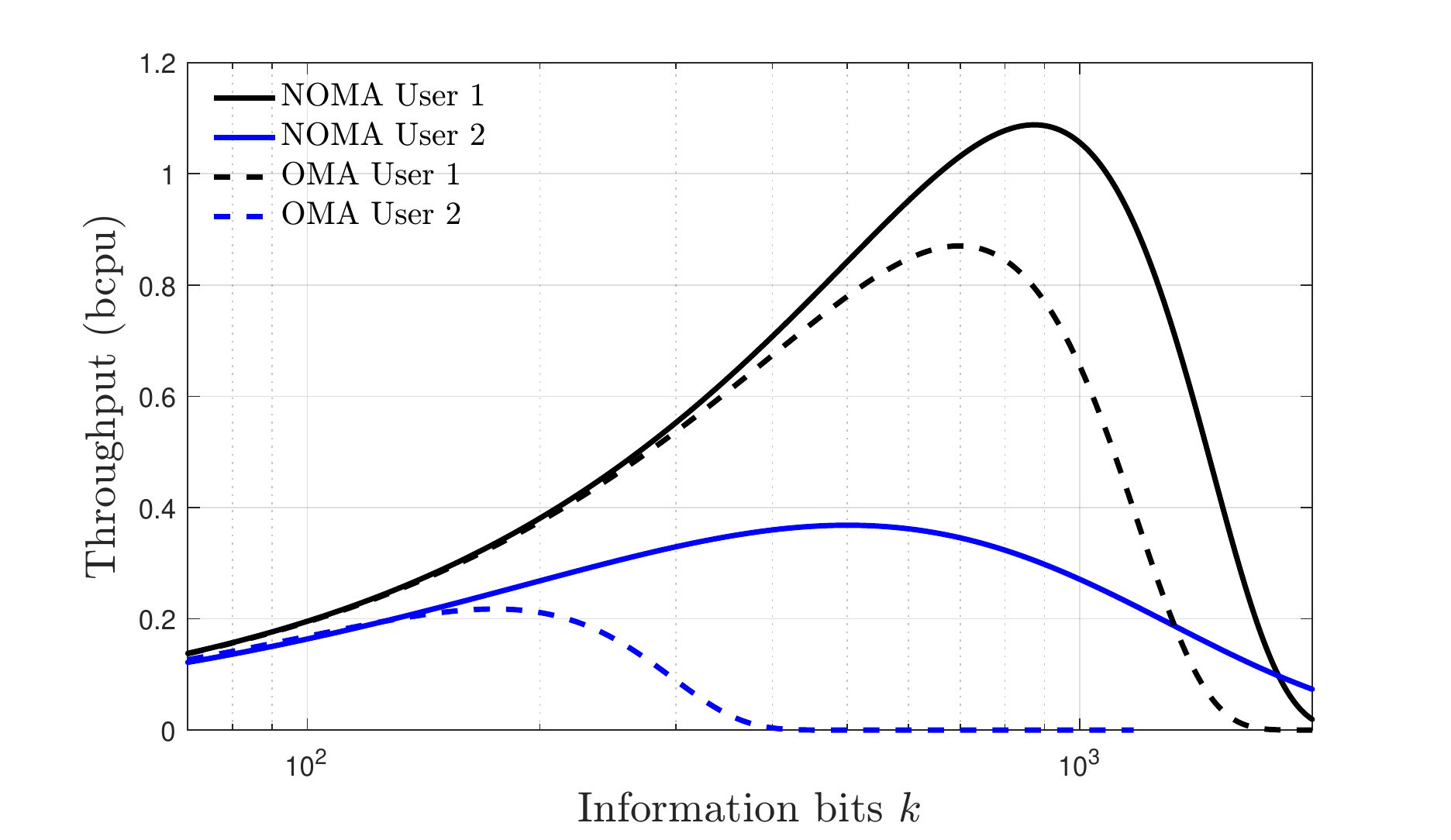}
	\caption{Throughput as a function of the number of information bits in Rayleigh fading channel, considering $n= 500$ channel uses and $P_1 = P_2=10$ dB.}
	\label{fig:k_eps}
\end{figure}

Now, we compare the performance of NOMA and OMA in quasi-static Rayleigh fading channel. First, we evaluate the throughput as a function of the number of information bits $k$. The result is shown in Fig. \ref{fig:k_eps} for $n=500$ and $P_1=P_2=10$ dB. From the figure, we notice that the non-orthogonal scheme outperforms OMA in the case of both users. When the number of information bits is large enough, we notice that for both users NOMA achieves higher throughput than OMA. Again the optimum number of information bits $k$ which maximizes the throughput differs from NOMA to OMA. To decrease the outage for the user we can increase the transmit power and apply ARQ with $M$ transmissions or decrease the rate as will be shown below. 

\begin{figure}[!htb] 
	\centering
	\includegraphics[clip, trim=1cm 0cm 0cm 0cm,width=1.08\columnwidth]{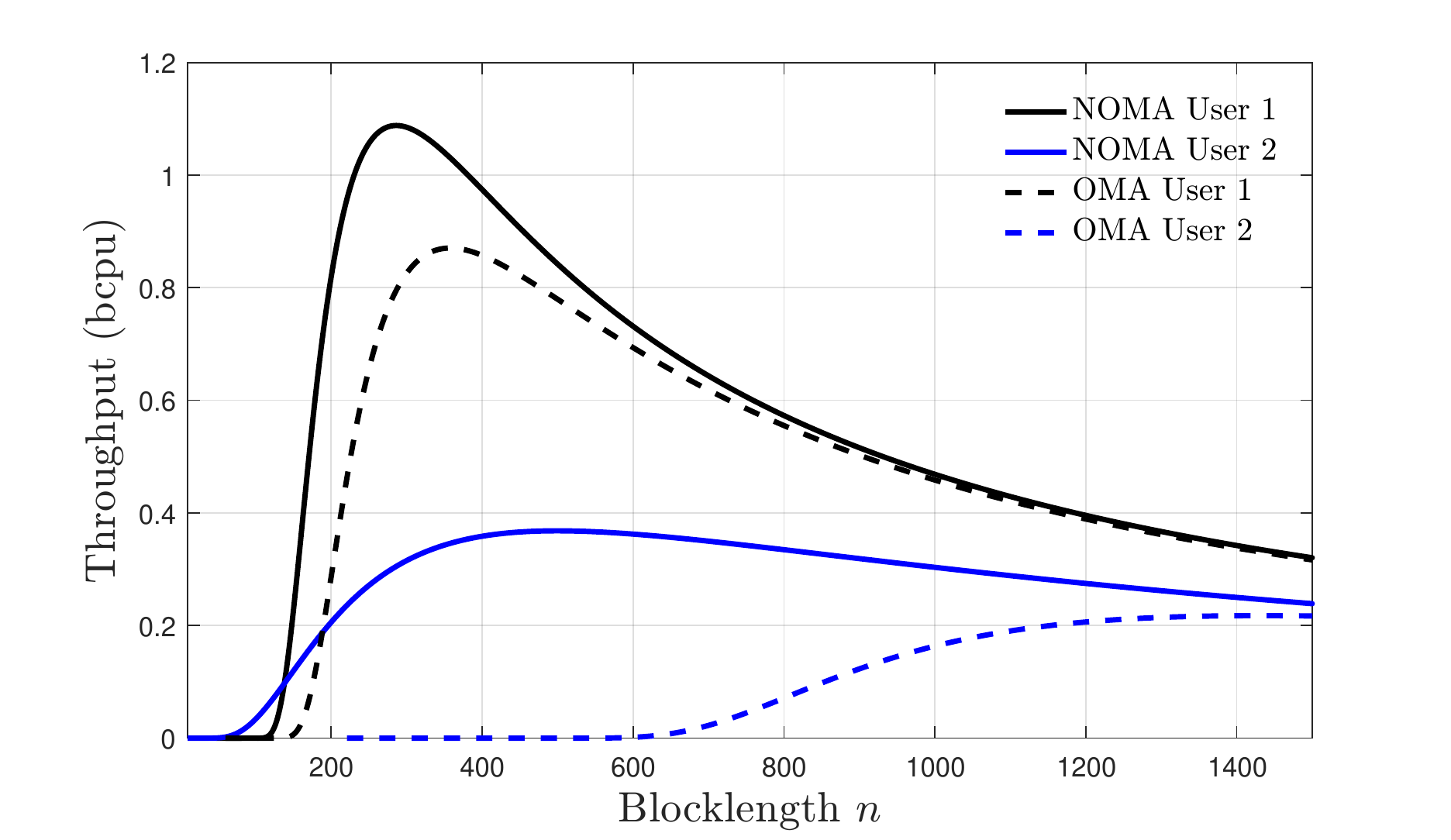}
	\caption{Throughput as a function of the number of channel uses in Rayleigh fading channel, considering $k= 500$ bits, $P_1 = P_2=10$ dB.}
	\label{fig:n_eps}
\end{figure}

Next, we analyze the throughput as a function of the number of channel uses. For this purpose, we fix $k=500$, and $\rho=10$ dB. From Fig. \ref{fig:n_eps}, we notice that in the short packet regime, in which we are operating, NOMA outperforms OMA for both users. For instance, when $n=500$ the throughput for the NOMA user 1 is more than the double when compare to the case of OMA. 

\begin{figure}[!htb] 
	\centering
	\includegraphics[clip, trim=1cm 0cm 0cm 0cm,width=1.08\columnwidth]{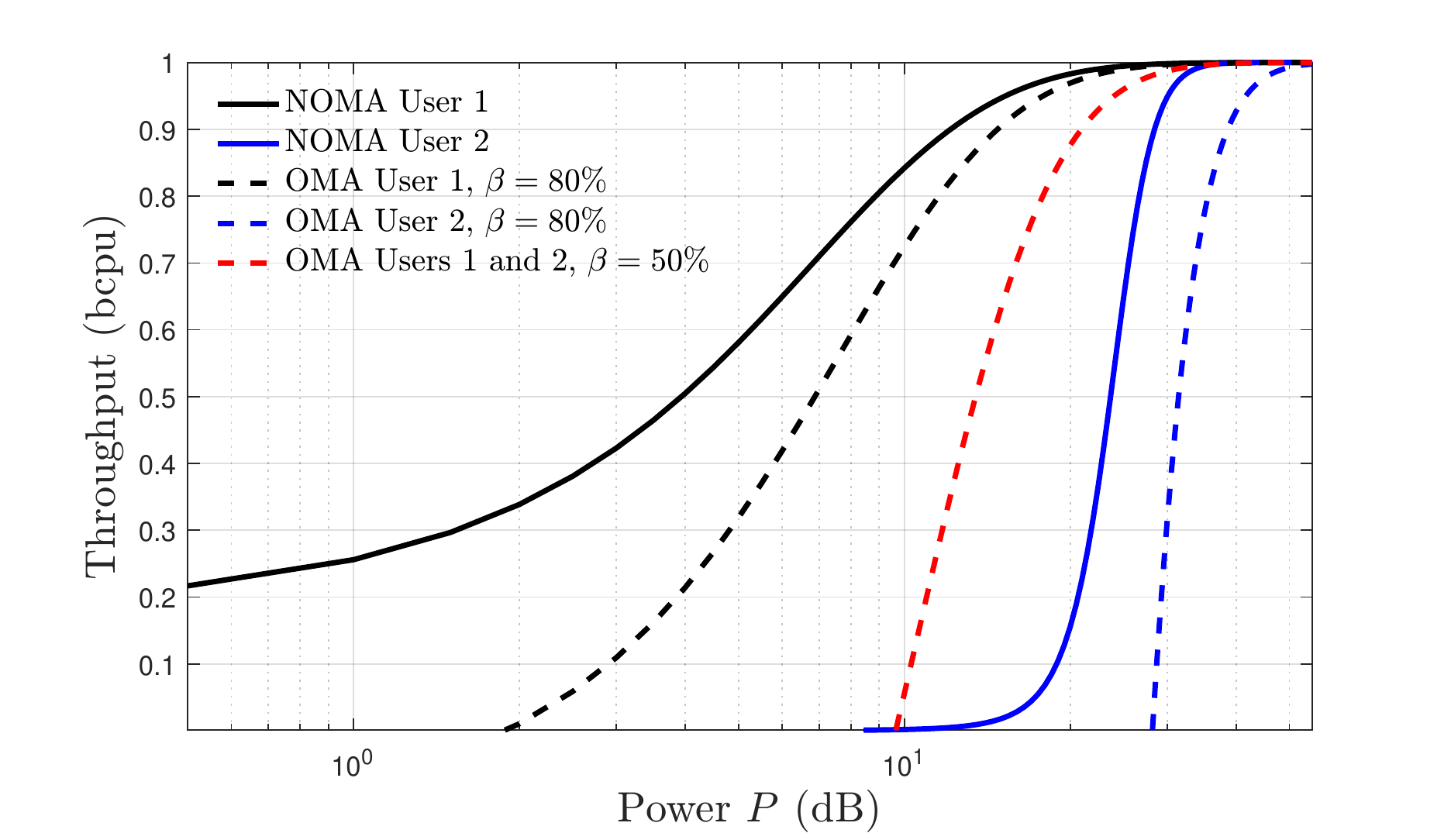}
	\caption{Transmitted power as a function of throughput $\eta$.}
	\label{fig:P_eps}
\end{figure}

Furthermore, we analyze the throughput as a function of the transmission power when $R=1$ bpcu and $n=k=500$. The result is shown in Fig. \ref{fig:P_eps}. Same as in Section \ref{crap}, for the OMA scheme we analyze the scenarios when highest priority is assigned to one user (i.e. $\beta=80\%$), or when both users are treated fairly (i.e. $\beta=50\%$). Again, we notice that NOMA maximizes the throughput while spending less power than OMA. However, when both users in OMA are treated equally, we notice that the power savings for the second user are higher. However, this comes at the cost of larger power expenditures, when compared to the user with highest priority in NOMA. 

\begin{figure}[!htb] 
	\centering
	\includegraphics[clip, trim= 0cm 7cm 0cm 6cm, width=1\textwidth]{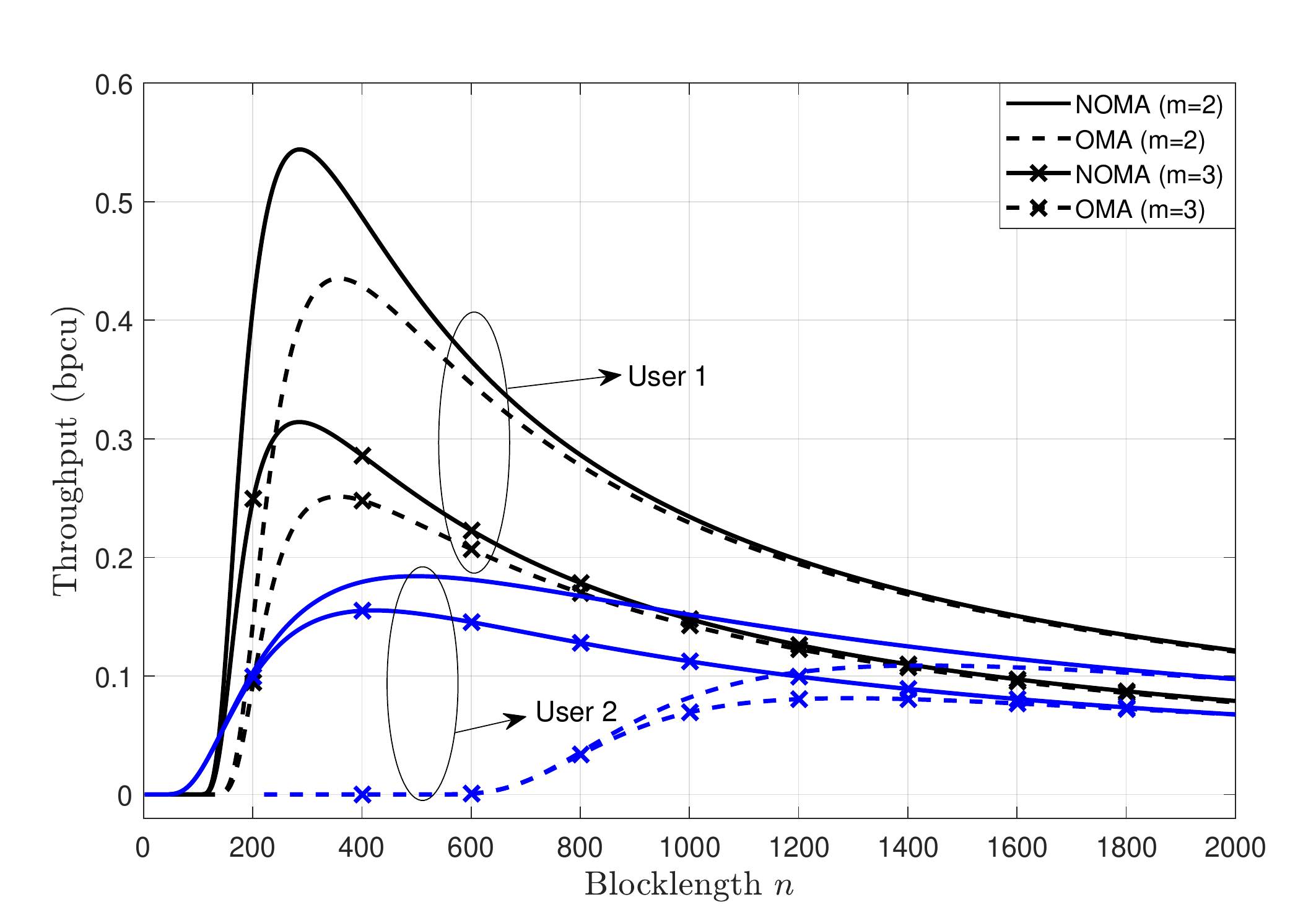}
	\caption{Throughput of ARQ in NOMA and OMA schemes as a function of blocklength $n$ for $P_1=P_2=10$ dB and $k=500$.}
	\label{Throughput2}
\end{figure}

In Fig. \ref{Throughput2}, the throughput of the ARQ protocol is plotted as a function of the blocklength $n$ in NOMA and OMA schemes for 2 and 3 transmissions. Here, we fix the system parameters as $P_1=P_2=10$ dB and $k=500$ and assume zero feedback delay $D=0$. The figure depicts the throughput declination introduced by applying type-I ARQ protocol which verifies our expectations for \eqref{eq:31}. The throughput even becomes worse when increasing the number of retransmissions $M$. However, NOMA still outperforms OMA for both users regardless of the number of retransmissions. It is noticed that the optimum blocklength for throughput maximization also changes with the number of transmissions. 

\begin{figure}[!htb] 
	\centering
	\includegraphics[clip, trim= 0cm 7cm 0cm 6cm, width=1\textwidth]{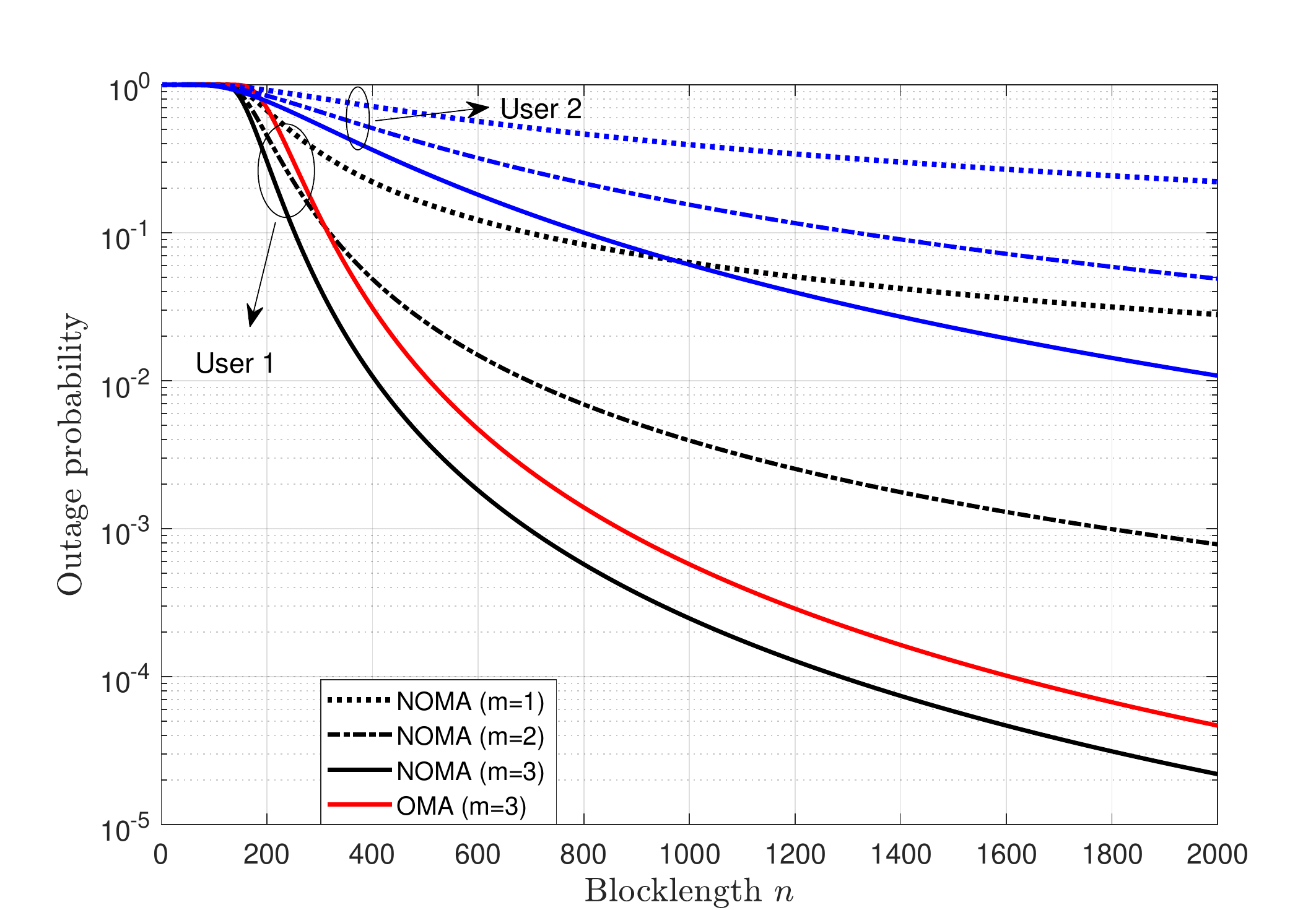}
	\caption{Outage probability of ARQ in NOMA and OMA schemes as a function of blocklength $n$ for $P_1=P_2=10$ dB and $k=500$.}
	\label{outage_plot}
\end{figure}

Fig. \ref{outage_plot} includes a plot for the outage probability as a function of the blocklength $n$ in NOMA and OMA schemes for $P_1=P_2=10$ dB and $k=500$. It illustrates the reliability enhancement which results from applying type-I ARQ protocol in terms of outage probability. It is obvious that increasing the number of retransmissions $M$ reduces the outage probability for both users and hence, boosts reliability. Thus, the relation between throughput and reliability is a trade-off which is subject to the design requirements. Moreover, the figure shows that for $M=3$, the outage probability of NOMA is lower than OMA for user 1 which confirms that the reliability of NOMA is higher than OMA. Here, the reliability is greater for user 1 and it increases for longer packets. Fig. \ref{Throughput2} and Fig. \ref{outage_plot} assure that despite the throughput decrease which occurs when applying type-I ARQ protocol, there is a significant improvement in reliability specially for NOMA. 

\begin{figure}[!htb] 
	\centering
	\includegraphics[width=1\textwidth]{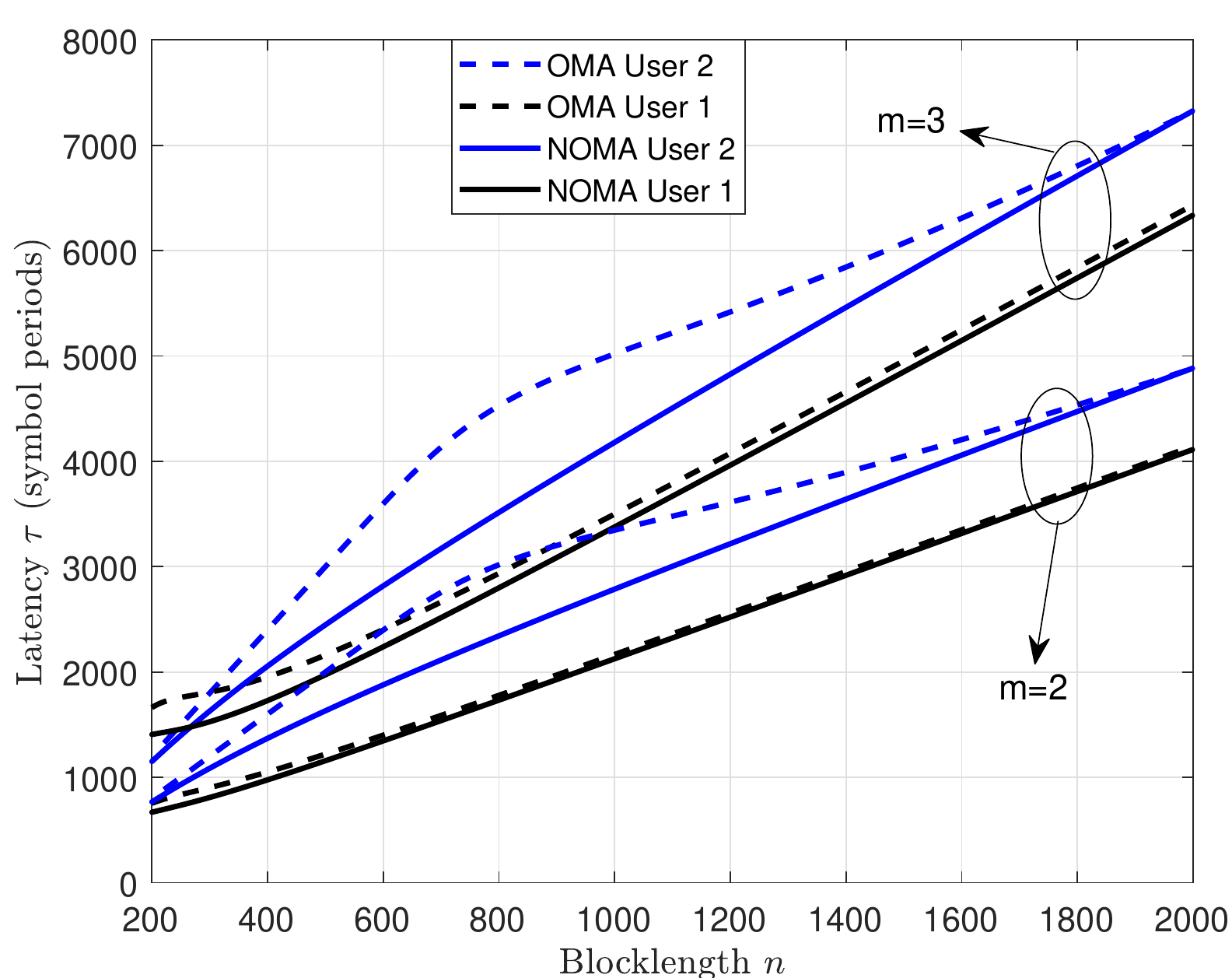}
	\caption{Latency of ARQ in NOMA and OMA schemes as a function of blocklength $n$ for $P_1=P_2=10$ dB and $k=500$.}
	\label{latency}
\end{figure}

Finally Fig. \ref{latency} elucidates the latency $\mathcal {T}$ as a function of the blocklength $n$ for both OMA and NOMA when applying ARQ. It is clear that enhancing reliability by increasing the number of retransmissions results in longer delays. However, the figure reveals that applying NOMA not only provides higher reliability as evinced in Fig. \ref{outage_plot}, but also renders lower latency specially for mid range blocklength ($\approx 800-1000$ symbol periods). Hence, NOMA matches the intuition of URLLC more than OMA in future 5G networks .


\section{Conclusions}\label{con}

In this work, we compared the performance of NOMA and OMA as two different multiple access methods in the finite block-length regime. First, we derived the outage expressions for NOMA and OMA scenarios in the finite blocklength regime. In our analysis, we considered AWGN and quasi-static Rayleigh fading channel models. The results showed that even for short packets, NOMA allows transmission with higher rate and throughput than OMA and thus, NOMA clearly outperforms OMA. Furthermore, we have observed that NOMA provides larger power gains when compared to OMA. We analyzed the performance of type-I ARQ protocol and quantized the throughput declination when increasing the number of retransmissions to improve reliability in terms of outage probability. The results revealed that although the throughput becomes poor for higher number of transmissions, applying ARQ significantly enhances the reliability of NOMA scheme which is still better than OMA in this case. Moreover, the latency introduced by ARQ is reduced by applying NOMA rather than OMA. As future work, we intend to analyze the performance of these schemes in massive Machine Type Communication (m-MTC) setups and introduce scheduling algorithms that can be deployed when the number of machines is large. Specifically it would be of high interest to evaluate the reliability and latency levels that can be meet by deploying such schemes.  
\vspace{-0mm}
\section{Acknowledgments}

This work is partially supported by Aka Project SAFE (Grant no. 303532), and by Finnish Funding Agency for Technology and Innovation (Tekes), Bittium Wireless, Keysight Technologies Finland, Kyynel, MediaTek Wireless, and Nokia Solutions and Networks.

\bibliography{mybibfile}

\end{document}